\documentclass{article}%
\usepackage{amsfonts}
\usepackage{amsmath}
\usepackage{amssymb}
\usepackage{graphicx}%
\providecommand{\U}[1]{\protect\rule{.1in}{.1in}}
\begin{document}

\title{Changes of Variables and the Renormalization Group\thanks{This work
\cite{Caticha 1984a} first appeared as Caltech preprint CALT-68-1099. It was
supported in part by the Department of Energy under contract DEAC
03-81-ER0050. }}
\author{Ariel Caticha\thanks{Current address: Department of Physics, University at
Albany--SUNY, Albany, NY 12222, USA}\\California Institute of Technology, Pasadena CA 91125}
\date{}
\maketitle

\begin{abstract}
A class of exact infinitesimal renormalization group transformations is
proposed and studied. These transformations are pure changes of variables
(\emph{i.e.}, no integration or elimination of some degrees of freedom is
required) such that a saddle point approximation is more accurate, becoming,
in some cases asymptotically exact as the transformations are iterated. The
formalism provides a simplified and unified approach to several known
renormalization groups. It also suggests some new ways in which
renormalization group methods might successfully be applied. In particular, an
exact gauge covariant renormalization group transformation is constructed.
Solutions for a scalar field theory are obtained both as an expansion in
$\varepsilon=4-d$ and as an expansion in a single coupling constant.

\end{abstract}

\section{Introduction}

For quite some time now it has been apparent that the success of the
renormalization group (RG) methods (see \emph{e.g.}, \cite{Caticha 1984a} and
references therein) in dealing with problems involving many degrees of freedom
is connected to an appropriate choice of variables. That is, the various RG's
provide systematic ways to focus one's attention on the degrees of freedom
that are most important in the problem under consideration. For example, in
Wilson's approach to the problem of critical phenomena \cite{Wilson Kogut
1974} it is recognized that short wavelength degrees of freedom are not
interesting in themselves, but only indirectly through the effective
interactions they induce between the experimentally accessible long wavelength
degrees of freedom. The strategy is to eliminate the short wavelengths in a
series of steps.\ One starts by eliminating the shortest ones first, then
slightly longer ones, and so on, gradually working one's way towards an
effective Lagrangian which contains only the relevant degrees of freedom.

In RG's such as the original Gell-Mann and Low RG \cite{Gell-Mann Low 1954}
the appropriate choice of variables was achieved in quite a different way. The
point is that while all wavelengths contribute to a loop integration in a
given Feynman diagram, the actual relative contribution of the short versus
the long wavelengths depends on the renormalization scale chosen. The freedom
to change the renormalization scale thus allows us to emphasize some degrees
of freedom over others and therefore to improve the perturbative calculation.

The two approaches above are sufficiently different that in spite of yielding
the same results when applied to a given problem the connection between the
two has been a matter of some confusion.

The Wilson RG transformation involves not only an elimination of some degrees
of freedom but also an explicit change of variables. Some consequences of this
fact appeared in works by Jona-Lasinio \cite{Jona-Lasinio 1973} and Wegner
\cite{Wegner 1974}. These authors were concerned with the possibility of
choosing more general RG transformations and showing that physically
significant quantities such as critical exponents are independent of such a
choice. Thus, Jona-Lasinio defines generalized renormalization transformations
as all those that leave the effective action $\Gamma$ invariant in value. It
seems unlikely that one can be more general than that, but this evades the
important issue of which transformations are useful. Wegner goes further. He
recognizes that transformations can be made in a rather general way and makes
the essential remark that elimination of degrees of freedom is not a necessary
step since some changes of variables effectively accomplish such an
elimination. He then goes on to exhibit explicitly the transformation that
generates Wilson's incomplete-integration RG \cite{Wilson Kogut 1974} and to
conjecture that useful transformations would involve some kind of nonlinearity
perhaps through some unspecified dependence on the Hamiltonian.

In this work \cite{Caticha 1984a} I study a class of exact infinitesimal RG
transformations for field theories in continuum space that are pure changes of
variables, \emph{i.e.}, no additional elimination or integration of certain
degrees of freedom is required. To isolate the minimal structure a change of
variables needs to include in order to actually accomplish this, I formulate
in section 2 three exact RG's in differential form. These are equivalent
though simplified versions of the sharp-cutoff RG of Wegner and Houghton
\cite{Wegner Houghton 1973}, of the incomplete integration RG of Wilson
\cite{Wilson Kogut 1974}, and of the hard-soft splitting RG \cite{Wilson
1976}-\cite{Shalloway 1979}.

In section 3 the required change of variables is obtained as well as the RG
equations both in functional form and as an infinite set of
integro-differential equations. The reason why this class of RG's is useful is
immediately apparent: the changes of variables are such that a classical or
saddle-point approximation in the new variables is more accurate. No mention
is made of the question of long versus short wavelengths; this is important.
On iterating the RG transformations (\emph{i.e.}, on solving the equations for
the RG evolution of the action or of the Hamiltonian) the classical
approximation becomes better and better approaching the exact result. Since
these RG equations are much simpler than other sets of equations that need to
be tackled in order to solve quantum field theories (\emph{e.g.}, the
Schwinger-Dyson equations), it suggests that this is a promising way (as is
the case with other RG's) to leap beyond the limitations of perturbation theory.

A further fortunate feature is related to the possibility of applying these
RG's to gauge theories. The original motivation for undertaking this study was
to construct an exact gauge covariant RG transformation. Such a transformation
would allow one to impose the stringent constraints of manifest gauge
invariance on the RG-evolved action and perhaps obtain a non-perturbative
solution of the RG equations. An analogous program has been partially carried
out by Baker, Ball, and Zachariasen (see \emph{e.g.}, \cite{Baker Ball
Zachariasen 1983} and references therein).

Once the structure of changes of variables that are also RG transformations is
identified the actual construction of gauge covariant transformations is
trivial. The application to gauge theories is discussed in another publication
\cite{Caticha 1984b}. The calculation of Green's functions is considered in
section 4. As with most manipulations with path-integrals the level of
mathematical rigor is fairly low. Changes of variables occasionally produce
surprises in that the new Lagrangian differs from the one that would be
naively obtained. In some situations the additional terms can be cast in the
form of an extra potential of order $\hbar^{2}$, in other situations they can
be traced to nontrivial Jacobian factors and they generate anomalies (see
\emph{e.g.}, \cite{Gervais Jevicki 1976}\cite{Fujikawa 1979} and references
therein). Two explicit solutions of the RG equations in section 5 serve as a
check that in our case no such surprises occur. The first solution is an
expansion in $\varepsilon=4-d$ \cite{Wilson Kogut 1974}\cite{Shukla Green
1974}, the second is an expansion in a single coupling constant. Both give the
same results, but they represent different viewpoints. The former emphasizes
analyticity, the latter is closer in spirit to the Gell-Mann and Low spirit.
Finally, the conclusions and some comments appear in section 6. Some of the
details of the calculations and a pedagogical example, a scalar field theory
in zero dimensions (a single integral) are discussed in the appendices.

\paragraph*{Note added:}

In the many years since this paper \cite{Caticha 1984a} was written a
considerable amount of research has been carried out on the subject which is
now variously known as the exact RG, the functional RG, and the
non-perturbative RG. Much of the material presented here has been
independently rediscovered, and there have been important developments that go
far beyond the original scope of this paper. Prominent among the latter is the
work by J. Polchinski \cite{Polchinski 1984} where exact RG methods were
developed as a method to prove renormalizability. The computational power of
the exact RG has been extended in the work by C. Wetterich \cite{Wetterich
1991} and coauthors --- the effective average action method --- including its
application to Yang-Mills theory \cite{Reuter Wetterich 1994} and gravity
\cite{Reuter 1998}. Wegner's original insight of the RG transformation as a
change of variables has been considerably expanded by T. R. Morris and
co-workers \cite{Latorre Morris 2000}\cite{Morris Preston 2016}. (For
additional references see the excellent reviews \cite{Morris 1994}-\cite{Nagy
2012}.) Despite such extensive literature the point of view presented here may
still have some pedagogical value since some of our results --- the
interpretation of the RG as a change of variables that systematically improves
a saddle point approximation, the calculation of the RG\ flow of Green's
functions, and the toy model of a field theory in zero dimensions --- do not
seem to have yet appeared in print.

\section{Three exact differential RG's}

\subsection{Sharp-Cutoff RG}

Here I present a modified version of the sharp-cutoff RG derived by Wegner and
Houghton \cite{Wegner Houghton 1973}. Consider the Green's function generating
functional in Euclidean space,
\begin{equation}
Z=\int D\phi\,\exp-S_{\tau}[\phi]~. \label{Z}%
\end{equation}
For the moment I will not be concerned with coupling the field $\phi$ to
external sources. This problem will be addressed in section 4.

Suppose field components with momenta larger than a certain cutoff
$\Lambda_{\tau}=\Lambda e^{-\tau}$ have been integrated out, \emph{i.e.},
\begin{equation}
\phi(q)=0\quad\text{for}\quad q>\Lambda_{\tau}~.
\end{equation}
This implies that the effective action $S_{\tau}$ consists not just of the
simple interactions contained in the bare action $S=S_{-\infty}$ but rather of
an infinite number of arbitrarily complicated interactions. Our problem is to
study how the action evolves when the cutoff is slightly decreased to
$\Lambda_{\tau+\delta\tau}$. Suppose we separate out the field components with
momenta in the thin shell between $\Lambda_{\tau+\delta\tau}$ and
$\Lambda_{\tau}$,
\begin{equation}
\phi(x)\rightarrow\phi(x)+\sigma(x)~,
\end{equation}
where on the right hand side $\phi(q)=0\ $for$\ q>\Lambda_{\tau+\delta\tau}$
and $\sigma(q)=0$ for $q$ outside the thin shell of thickness $\Lambda_{\tau
}\delta\tau$. On integrating out the $\sigma$ fields the new action will be
given by
\begin{equation}
\exp-S_{\tau+\delta\tau}[\phi]=\int D\sigma\,\exp-S_{\tau}[\phi+\sigma]~.
\label{new action}%
\end{equation}
The integration is performed perturbatively in three steps: first expand
$S_{\tau}[\phi+\sigma]$ in a power series in $\sigma$; second, isolate a
convenient $\sigma$-field propagator; and third, treat the remaining
$\sigma^{p}$ vertices ($p\geq1$) as a perturbation. This procedure, carried
out in detail in Appendix A shows that to first order in the shell thickness
only diagrams with one internal $\sigma$-field line contribute. The result is
\begin{equation}
S_{\tau+\delta\tau}[\phi]-S_{\tau}[\phi]=\int d^{d}q\,(2\pi)^{d}\Delta_{\tau
}(q)\left[  \frac{\delta^{2}S_{\tau}}{\delta\phi(q)\delta\phi(-q)}%
-\frac{\delta S_{\tau}}{\delta\phi(q)}\frac{\delta S_{\tau}}{\delta\phi
(-q)}\right]  \,, \label{sharp cutoff a}%
\end{equation}
where $\Delta_{\tau}(q)$ is a convenient $\sigma$ propagator which vanishes
outside the shell. Alternatively,
\begin{equation}
S_{\tau+\delta\tau}[\phi]-S_{\tau}[\phi]=(2\pi)^{d}\Lambda_{\tau}^{d-2}%
\delta\tau\int d\Omega_{d}\,\left[  \frac{\delta^{2}S_{\tau}}{\delta
\phi(q)\delta\phi(-q)}-\frac{\delta S_{\tau}}{\delta\phi(q)}\frac{\delta
S_{\tau}}{\delta\phi(-q)}\right]  \,, \label{sharp cutoff b}%
\end{equation}
where now $q^{2}=$ $\Lambda_{\tau}^{2}$ and $d\Omega_{d}\,$\ is the element of
solid angle in $d$ dimensions.

To obtain RG equations an additional dilatation change of variables is
required. This is a trivial step which will be addressed later in section 3.

The basic improvement of eqs.(\ref{sharp cutoff a}) and (\ref{sharp cutoff b})
over those of Wegner and Houghton is that their equation include all diagrams
with one internal $\sigma$-field loop (\emph{i.e.}, many internal $\sigma$
propagators) while ours include only the much smaller set of diagrams with
only one $\sigma$-field propagator.

\subsection{Incomplete-integration RG}

In order to avoid the unphysical difficulties introduced by the discontinuous
cutoff considered in the last section, Wilson \cite{Wilson Kogut 1974}
introduced the concept of incomplete integration designed to achieve a smooth
interpolation between those degrees of freedom that have been integrated out
and those that have not.

The idea is implemented through the introduction of an auxiliary functional
$\delta_{\alpha}[\phi]$. In the case of an ordinary single integral
$\delta_{\alpha}(x)$ is a function such that as $\alpha$ goes from $0$ to
$\infty$, the function
\[
z_{\alpha}(x)=\int dy\,z_{0}(y)\delta_{\alpha}(y-x)
\]
smoothly interpolates between the integrand $z_{0}(y)$ and the integral
\[
z_{\infty}(x)=\int dy\,z_{0}(y)~.
\]
All that is required is that
\[
\delta_{\alpha}(x)\rightarrow\left\{
\begin{array}
[c]{ccc}%
\delta(x) & \text{for} & \alpha\rightarrow0~,\\
1 & \text{for} & \alpha\rightarrow\infty~.
\end{array}
\right.
\]
Wilson's choice for $\delta_{\alpha}$ was the Green's function of a certain
differential equation. It is perhaps simpler to use a Gaussian,
\[
\delta_{\alpha}(x)=\left(  \frac{1}{4\pi\alpha}+1\right)  ^{1/2}\exp
-\frac{x^{2}}{4\alpha}~.
\]
The case of a single integral is pursued further in appendix B. now we return
to the functional integral problem. We let $\alpha=\alpha(q,\tau)$ and
introduce
\[
\text{const.}=\int D\phi\,\delta_{\alpha}[\phi-\Phi]
\]
into the functional integral
\[
Z=\int D\Phi\,\exp-S[\Phi]~.
\]
The result is
\[
Z=\int D\phi\,\exp-S_{\tau}[\phi]~,
\]
where
\begin{align}
\exp-S_{\tau}[\phi]  &  =\int D\Phi\,\delta_{\alpha}[\phi-\Phi]\exp
-S[\Phi]\nonumber\\
&  =\int D\Phi\,\exp-\left(  S[\Phi]+\int d\tilde{q}\frac{|\phi(q)-\Phi
(q)|^{2}}{4\alpha(q,\tau)}\right)  ~, \label{smooth cutoff a}%
\end{align}
where we drop an unimportant field-independent factor. In
eq.(\ref{smooth cutoff a}) and in the following we adopt the notation
\[
d\tilde{q}=\frac{d^{d}q}{(2\pi)^{d}}\quad\text{and}\quad\tilde{\delta
}(q)=(2\pi)^{d}\delta^{d}(q)~.
\]

The conventional usage is to choose $\alpha(q,\tau)$ such that $S_{\tau}%
[\phi]$ describes modes with an effective cutoff $\Lambda_{\tau}=\Lambda
e^{-\tau}$ which means that $\alpha(q,\tau)$ is very large for $q\gg
\Lambda_{\tau}$ and very small for $q\ll\Lambda_{\tau}$. A convenient, but by
no means obligatory choice is one in which the mode $\phi(qe^{-\delta\tau})$
in $S_{\tau+\delta\tau}$ is integrated out to the same extent as the mode
$\phi(q)$ in $S_{\tau}$, \emph{i.e.},
\[
\alpha(qe^{-\delta\tau},\tau+\delta\tau)=\alpha(q,\tau)~.
\]
This implies that $\alpha$ depends on $q$ and $\tau$ only through the
combination $qe^{\tau}$. Let%
\begin{equation}
\alpha_{\tau}(q)=\alpha(q/\Lambda_{\tau})~.
\end{equation}

The functional integral (\ref{smooth cutoff a}) can be transformed into a
functional differential equation,
\begin{equation}
\frac{d}{d\tau}S_{\tau}=\int d^{d}q\,(2\pi)^{d}\dot{\alpha}_{\tau}(q)\left[
\frac{\delta^{2}S_{\tau}}{\delta\phi(q)\delta\phi(-q)}-\frac{\delta S_{\tau}%
}{\delta\phi(q)}\frac{\delta S_{\tau}}{\delta\phi(-q)}\right]
\label{smooth cutoff b}%
\end{equation}
where $\dot{\alpha}_{\tau}=d\alpha_{\tau}/d\tau$. This equation is obtained
noticing that differentiation of (\ref{smooth cutoff a}) with respect to
$\tau$ brings down factors of $(\phi-\Phi)$ on the right hand side that may
also be brought through functional differentiation with respect to $\phi$.
Comparison of eq.(\ref{smooth cutoff b}) with the remarkably similar
eq.(\ref{sharp cutoff a}) shows that $\dot{\alpha}_{\tau}d\tau$ is playing the
role of a propagator.

Again, the full RG equations require an additional dilation which we postpone
until section 3.

\subsection{Hard-Soft Splitting RG}

Another method which allows elimination of short wavelength degrees of freedom
was suggested by Wilson \cite{Wilson 1976}. It consists of splitting the
propagator into two pieces, one contributes dominantly for high momenta while
the other does so for low momenta,
\begin{equation}
\frac{1}{p^{2}}=\frac{1}{p^{2}+\mu^{2}}+\frac{\mu^{2}}{p^{2}(p^{2}+\mu^{2})}~.
\label{hard soft a}%
\end{equation}
The idea was to take advantage of the UV asymptotic freedom of Yang-Mills
theories to integrate out the hard components in renormalized perturbation
theory to generate an effective action for the soft components which could be
treated using techniques better suited to the strong coupling regime.

The method was further developed by Lowenstein and Mitter \cite{Lowenstein
Mitter 1977} and by Mitter and Valent \cite{Mitter Valent 1977}, and applied
to the weak coupling regime of quantum chromodynamics by Shalloway
\cite{Shalloway 1979}. In this section we formulate it in a simple way which
allows immediate comparison with the RG's described in the previous sections.

The actual splitting of hard and soft components is accomplished by
introducing
\[
\text{const.}=\int D\chi\,\exp-\int dx\frac{1}{2}\mu^{2}\chi^{2}%
\]
into the path integral
\[
Z=\int D\Phi\,\exp-S[\Phi]\quad\text{where}\quad S[\Phi]=\int dx\left[
\frac{1}{2}\partial\Phi\partial\Phi+V(\Phi)\right]  \,,
\]
and then making the change of variables $(\Phi,\chi)\rightarrow(\phi,\phi
_{h})$ where
\[
\Phi=\phi+\phi_{h}\quad\text{and}\quad\chi=\phi_{h}+\frac{\partial^{2}}%
{\mu^{2}}\phi~.
\]
The result is
\[
Z=\int D\phi D\phi_{h}\exp\int dx\left[  \frac{1}{2}\phi\partial^{2}\left(
1-\frac{\partial^{2}}{\mu^{2}}\right)  \phi+\frac{1}{2}\phi_{h}(\partial
^{2}-\mu^{2})\phi_{h}-V(\phi+\phi_{h})\right]  \,,
\]
where the hard-soft separation (\ref{hard soft a}) is explicit. Integrating
over $\phi_{h}$ leads once more to
\[
Z=\int D\phi\,\exp-S_{\tau}[\phi]~,
\]
where%
\[
\exp-S_{\tau}[\phi]=\int D\phi_{h}\,\exp-\left[  S[\phi+\phi_{h}]+\int
dx\frac{1}{2}\mu^{2}\left(  \phi_{h}+\frac{\partial^{2}}{\mu^{2}}\phi\right)
^{2}\right]  ~.
\]

We wish to study the evolution of $S_{\tau}$ under changes of $\tau$. We are
taking $\mu=\mu(\tau)$. It is convenient to shift integration variables back
to $\Phi$. Then%
\[
\exp-S_{\tau}[\phi]=\int D\Phi\,\exp-\left[  S[\Phi]+\int d\tilde{q}\frac
{1}{2}\mu^{2}\left(  \rho\phi(q)-\Phi(q)\right)  ^{2}\right]  ~
\]
where $\rho=1+q^{2}/\mu^{2}$. This equation, which is very similar to
(\ref{smooth cutoff a}), can also be transformed into a functional
differential equation,
\begin{equation}
\frac{dS_{\tau}}{d\tau}=\int\frac{d^{d}q}{2\rho^{2}}\left[  \,(2\pi)^{d}%
\frac{d\mu^{-2}}{d\tau}\left(  \frac{\delta^{2}S_{\tau}}{\delta\phi
(q)\delta\phi(-q)}-\frac{\delta S_{\tau}}{\delta\phi(q)}\frac{\delta S_{\tau}%
}{\delta\phi(-q)}\right)  +\frac{d\rho^{2}}{d\tau}\phi(q)\frac{\delta S_{\tau
}}{\delta\phi(q)}\right]  \,. \label{hard soft b}%
\end{equation}
This differs from eq.(\ref{sharp cutoff a}) and \ref{smooth cutoff a} only in
the last term, which by now one might suspect is not essential.

Incidentally, one could consider situations where $\mu$ and $\rho$ have
momentum dependencies other than those assumed above, in particular one could
choose
\begin{equation}
\,(2\pi)^{d}\mu^{-2}=\rho^{2}+c
\end{equation}
where $c$ is independent of $\tau$. Then eq.(\ref{hard soft b}) becomes
identical with the original incomplete-integration equation of Wilson
[eq.(11.8) of ref.\cite{Wilson Kogut 1974}].

The important conclusion to be drawn is that the various examples of exact
RG's considered above are characterized by a certain common feature, which we
might guess is what makes them useful RG's in the first place. The variations
are presumably inessential in principle, although in practice they may be
important. For example, the sharp-cutoff RG is definitely more inconvenient to
calculate with.

\section{The RG as a change of variables}

Functional integrals are a particularly convenient way to formulate quantum
field theories not just because they readily allow for perturbative expansions
but also because the implications of invariances and of the changes induced by
transformations of the dynamical variables can be easily studied. This feature
has been found particularly useful in \ the case of non-Abelian gauge
transformations. In this section we consider infinitesimal variable changes
that reproduce the exact RG transformations described previously.

Let us go back to eq.(\ref{Z}) and investigate the changes in the action
$S_{\tau}$ induced by the variable transformation
\begin{equation}
\phi(q)\rightarrow\phi^{\prime}(q)=\phi(q)+\delta\tau\,\eta_{\tau}[\phi,q]~,
\label{ch var a}%
\end{equation}
where $\eta_{\tau}[\phi,q]$ is some sufficiently well-behaved functional of
$\phi$ and a function of $q$. Taking into account the Jacobian of this
transformation eq.(\ref{Z}) becomes,
\begin{align*}
Z  &  =\int D\phi\,\left[  1+\delta\tau\int dq\frac{\delta\eta_{\tau}[\phi
,q]}{\delta\phi(q)}\right]  \exp-\left[  S_{\tau}[\phi]+\delta\tau\int
dq\frac{\delta S_{\tau}}{\delta\phi(q)}\eta_{\tau}[\phi,q]\right] \\
&  =\int D\phi\,\exp-S_{\tau+\delta\tau}[\phi]~,
\end{align*}
\ where%
\begin{equation}
S_{\tau+\delta\tau}[\phi]=S_{\tau}[\phi]+\delta\tau\int dq\left[  \frac{\delta
S_{\tau}}{\delta\phi(q)}\eta_{\tau}[\phi,q]-\frac{\delta\eta_{\tau}[\phi
,q]}{\delta\phi(q)}\right]  ~. \label{ch var b}%
\end{equation}
Suppose one chooses
\begin{equation}
\eta_{\tau}[\phi,q]=-(2\pi)^{d}\dot{\alpha}_{\tau}(q)\frac{\delta S_{\tau}%
}{\delta\phi(-q)}~, \label{ch var c}%
\end{equation}
then eq.(\ref{ch var b}) becomes identical to the Gaussian
incomplete-integration RG. More generally, if one also includes an inessential
rescaling of the field,
\[
\eta_{\tau}[\phi,q]=-(2\pi)^{d}\dot{\alpha}_{\tau}(q)\frac{\delta S_{\tau}%
}{\delta\phi(-q)}+\zeta_{\tau}(q)\phi(q)~,
\]
one obtains an equation of the form of eq.(\ref{hard soft b}).

The conclusion is that transformations of the form of eq.(\ref{ch var c}) are
RG transformations.

Furthermore, one can see why they are useful transformations. A field
configuration that is a solution of the classical equation of motion $\delta
S_{\tau}/\delta\phi=0$ will not be affected by (\ref{ch var c}). Any other
configuration will flow with $\tau$ until it becomes a classical solution
(\emph{i.e.}, a stationary point), then it ceases to flow. As $\tau
\rightarrow\infty$ a situation is approached in which all field configurations
are classical solutions, \emph{i.e.}, $\delta S_{\tau}/\delta\phi=0$ for all
$\phi$. The action approaches a constant.

In appendix B a toy example in zero spacetime dimensions, an ordinary
integral, is worked out. It allows one to see very clearly what is happening.
The changes of variables are such that a \textquotedblleft
classical\textquotedblright\ approximation, \emph{i.e.}, a steepest descent
approximation becomes better and better as $\tau$ increases, approaching the
exact result as $\tau\rightarrow\infty$. The reason the approximation is
improved is not that the integrand becomes steeper as one might at first
guess, but rather that it approaches a Gaussian for which the steepest descent
method is exact. The fact that this Gaussian is increasingly flatter (the
action becomes a constant) is not a serious obstacle. It merely requires us to
calculate the integral before the limit $\tau=0$ is reached.

Traditionally RG techniques have been applied to problems that exhibit some
kind of symmetry under changes of scale. In these cases it is convenient to
perform an additional change of variables in the form of a dilatation.
Consider a situation in which the effective cutoff is $\Lambda$. Under the
change
\[
\phi(q)\rightarrow\phi(q)+\delta\tau\,\eta_{0}[\phi,q]\quad\text{where}%
\quad\eta_{0}[\phi,q]=\left.  \eta_{\tau}[\phi,q]\right\vert _{\tau=0}~
\]
the effective cutoff is changed to $\Lambda e^{-\delta\tau}$. The scaling
transformation $q\rightarrow qe^{-\delta\tau}$ then guarantees that the new
momenta will span the same range $(0,\Lambda)$ as before. Thus one takes
\[
\delta_{\text{dil}}\phi(q)=\delta\tau\left(  d-d_{\phi}+q\cdot\frac{\partial
}{\partial q}\right)  \phi(q)~,
\]
where the field scale dimension,
\[
d_{\phi}=\frac{d}{2}-1+\gamma_{\phi}~,
\]
includes an anomalous dimension term.

The full RG transformation is
\begin{equation}
\phi(q)\rightarrow\phi(q)+\delta\tau\,\eta_{0}[\phi,q]+\delta\tau\left(
d-d_{\phi}+q\cdot\frac{\partial}{\partial q}\right)  \phi(q)~,
\label{ch var d}%
\end{equation}
and the full RG equation is
\begin{align}
\frac{d}{d\tau}S_{\tau}  &  =\int d^{d}q\,(2\pi)^{d}\dot{\alpha}(q)\left[
\frac{\delta^{2}S_{\tau}}{\delta\phi(q)\delta\phi(-q)}-\frac{\delta S_{\tau}%
}{\delta\phi(q)}\frac{\delta S_{\tau}}{\delta\phi(-q)}\right] \nonumber\\
&  +\frac{\delta S_{\tau}}{\delta\phi(q)}\left(  d-d_{\phi}+q\cdot
\frac{\partial}{\partial q}\right)  \phi(q)~, \label{ch var e}%
\end{align}
where now $\dot{\alpha}=d\alpha/d\tau|_{\tau=0}$.

The functional equation (\ref{ch var e}) can be transformed into an infinite
set of integro-differential equations. Substituting an action of the general
form
\begin{equation}
S_{\tau}[\phi]=\sum_{n~\text{even}}^{\infty}\frac{1}{n!}\int d\tilde{q}%
_{1}\ldots d\tilde{q}_{n}\tilde{\delta}\left(
%TCIMACRO{\tsum \nolimits_{j}^{n}}%
%BeginExpansion
{\textstyle\sum\nolimits_{j}^{n}}
%EndExpansion
q_{j}\right)  u_{n}(q_{1}\ldots q_{n},\tau)\phi(q_{1})\ldots\phi(q_{n})
\label{ch var f}%
\end{equation}
into eq.(\ref{ch var e}) and equating the coefficients of terms of the same
degree in $\phi$ one obtains (omitting the $\tau$ dependence)%
\begin{gather}
\frac{\partial}{\partial\tau}u_{n}(q_{1}\ldots q_{n})=\int d\tilde{k}%
\,\dot{\alpha}(k)u_{n+2}(q_{1}\ldots q_{n},k,-k)\nonumber\\
+\sum_{m\geq2}^{n}\binom{n}{m-1}\frac{1}{n!}\sum_{\{q_{j}\}}\dot{\alpha}%
(k_{m})u_{m}(q_{1}\ldots q_{m-1},k_{m})u_{n-m+2}(q_{m}\ldots q_{n}%
,-k_{m})\nonumber\\
+\left[  d-nd_{\phi}-%
%TCIMACRO{\tsum \nolimits_{j-1}^{n}}%
%BeginExpansion
{\textstyle\sum\nolimits_{j-1}^{n}}
%EndExpansion
q_{j}\cdot\frac{\partial}{\partial q_{j}}\right]  u_{n}(q_{1}\ldots q_{n})~,
\label{ch var g}%
\end{gather}
where
\[
k_{m}=-\sum_{j=1}^{m-1}q_{j}=\sum_{j=m}^{n}q_{j}%
\]
and where $%
%TCIMACRO{\tsum \nolimits_{\{q_{j}\}}}%
%BeginExpansion
{\textstyle\sum\nolimits_{\{q_{j}\}}}
%EndExpansion
$ denotes a sum over all the permutations of the $q_{j}{}$'s.

Equations (\ref{ch var g}) can be given a simple graphical representation in
which the first term on the right hand side is a loop diagram with
$\dot{\alpha}(k)$ playing the role of the internal line propagator; the second
term is a tree diagram where $\dot{\alpha}(k_{m})$ is again the propagator for
the internal line. When a sharp-cutoff is employed, as discussed in section
2.2, the $\dot{\alpha}$'s do actually correspond to propagators in the
conventional sense.

\section{Green's Functions}

The calculation of Green's functions or of correlation functions brings us to
the problem of coupling the field $\phi$ to an external source. The study of
how Green's functions calculated from the bare action $S$ are related to those
calculated from $S_{\tau}$ can be done in a number of ways (see \emph{e.g.
}\cite{Wilson Kogut 1974}). We would like to address this question in the
spirit of the previous section, regarding the RG as an infinitesimal change of variables.

Consider the generating functional
\[
Z_{\tau}[j]=\int D\phi\,\exp\left(  -S_{\tau}[\phi]+\int j\phi\right)  ~.
\]
Performing a change of variables of the form of eq.(\ref{ch var b}) (for
simplicity we do not include the dilatation change of variables) we obtain,
\[
Z_{\tau}[j]=\int D\phi\,\left[  1-\delta\tau\int d^{d}q\,j(-q)\dot{\alpha
}_{\tau}(q)\frac{\delta S_{\tau}}{\delta\phi(-q)}\right]  \exp\left(
-S_{\tau}[\phi]+\int j\phi\right)  ~.
\]
But,
\[
0=\int D\phi\,\frac{\delta}{\delta\phi(-q)}e^{-S_{\tau}[\phi]+\int j\phi}=\int
D\phi\,\left[  \frac{j(q)}{(2\pi)^{d}}-\frac{\delta S_{\tau}}{\delta\phi
(-q)}\right]  e^{-S_{\tau}[\phi]+\int j\phi}~,
\]
and therefore%
\[
\frac{dZ_{\tau}}{d\tau}=\left[  \int d\tilde{q}\,j(-q)\dot{\alpha}_{\tau
}(q)j(q)\right]  Z_{\tau}~.
\]
Integrating in $\tau$ with the initial condition $S_{-\infty}=S$,
\emph{i.e.},
\[
Z_{-\infty}[j]=Z[j]=\int D\phi\,\exp\left(  -S[\phi]+\int j\phi\right)  ~,
\]
leads to
\[
Z[j]=\left[  \int d\tilde{q}\,j(-q)\alpha_{\tau}(q)j(q)\right]  Z_{\tau}[j]~,
\]
which exhibits the desired relation in a particularly simple form.

The generating functional of connected Green's functions, $W=-\log Z$, and the
corresponding $W_{\tau}$ are related by
\[
W[j]=W_{\tau}[j]+\int d\tilde{q}\,j(-q)\alpha_{\tau}(q)j(q)~.
\]
This shows that the connected $n$-point functions computed with $S_{\tau}$ are
identical with those computed with $S$ for $n\geq3$ while for $n=2$ they
differ in a rather trivial way. In this formulation it is then particularly
clear that physically significant quantities such as critical exponents or
$S$-matrix elements can be computed with either $Z$ or $Z_{\tau}$ and that
they are independent of the choice of $\alpha$, that is, independent of the
choice of the RG.

\section{Solutions}

Obtaining solutions to the RG equations (\ref{ch var g}) is a challenging
problem. In this section two conventional approximations are discussed, an
expansion in $\varepsilon=4-d$ and an expansion in a single coupling constant.
One motivation is to give us confidence that the expressions in the previous
sections are correct in spite of the lack of mathematical rigor employed in
their deduction. Another motivation is to compare two approximation schemes
which, although leading to differential equations of similar structure,
represent different viewpoints. Finally, yet a third motivation is to stress
the larger freedom of choice of the RG. This is important, not only because it
allows one to construct RG's in which the usual restriction of integrating
only over the short wavelength degrees of freedom is lifted, but also because
it will allow us to construct gauge covariant RG's.

A standard approach to solving eq.(\ref{ch var g}) consists of finding a fixed
point and studying the evolution of small perturbations about this fixed
point. One looks for a fixed point solution $S^{\ast}$ for which the vertex
functions $u_{n}^{\ast}$ do not depend on $\tau$ as an expansion in
$\varepsilon$,
\begin{align}
u_{2}^{\ast}  &  =V_{20}+V_{21}\varepsilon+V_{22}\varepsilon^{2}%
+\ldots\nonumber\\
u_{4}^{\ast}  &  =V_{41}\varepsilon+V_{42}\varepsilon^{2}+\ldots\nonumber\\
u_{6}^{\ast}  &  =V_{62}\varepsilon^{2}+\ldots\label{sol 1a}%
\end{align}
with the anomalous dimension given by
\begin{equation}
\gamma_{\phi}=\gamma_{1}\varepsilon+\gamma_{2}\varepsilon^{2}+\ldots
\label{sol 1b}%
\end{equation}
The crucial extra condition imposed on the solution and on the RG
transformation (\ref{ch var d}) (\emph{i.e.}, on the anomalous dimension
$\gamma_{\phi}$) is that the solution be analytic in the momenta. The need for
this condition can be vaguely argued as follows. Non-analyticity in momentum
space translates into long range of nonlocal interactions in position space
for which some features of critical behavior, like universality, are known not
to hold.

Details of these calculations, which are similar to those obtained in
\cite{Shukla Green 1974} for Wilson's incomplete-integration RG , can be found
in appendix C.

An alternative perturbative approach consists in expanding in a single
coupling constant $g(\tau)$ without referring to any fixed point. One looks
for a solution of the form
\begin{align}
u_{2}  &  =U_{20}+gU_{41}+g^{2}U_{42}+\ldots\nonumber\\
u_{4}  &  =\Lambda^{\varepsilon}(gV_{41}+g^{2}U_{42}+\ldots)\nonumber\\
u_{6}  &  =\Lambda^{2\varepsilon}(g^{2}U_{62}+\ldots)~, \label{sol 2a}%
\end{align}
with the anomalous dimension given by
\begin{equation}
\gamma_{\phi}=\gamma_{1}g+\gamma_{2}g^{2}+\ldots\label{sol 2b}%
\end{equation}
and $g(\tau)$ flowing according to
\begin{equation}
\frac{dg}{d\tau}=-\beta(g)=b_{1}g+b_{2}g^{2}+\ldots
\end{equation}
Factors of $\Lambda^{\varepsilon}$ have been made explicit so that various
$U$'s have the same dimensions they would have in $d=4$. The crucial extra
condition imposed on the solution and on the RG transformation in this
perturbative approach is that all $\tau$ dependence be contained in the single
function $g(\tau)$. This brings us somewhat closer to the spirit of the RG of
Gell-Mann and Low. The other functions are required not to depend on $\tau$
but could be non-analytic. These calculations are carried out in appendix D.

While none of the results obtained in those calculations are new the freedom
of the choice of the cutoff function $\dot{\alpha}(q)$ is explicit. In
particular, one is not required to integrate only short wavelengths, that is
$\dot{\alpha}(0)=0$, as for example in the usual choice $\dot{\alpha}%
(q)=q^{2}/\Lambda^{4}$. One can integrate the long wavelengths as well, for
example $\dot{\alpha}(q)=\Lambda^{-2}\exp(q^{2}/\Lambda^{2})$ or even
integrate all wavelengths simultaneously to the same extent by taking
$\dot{\alpha}=\Lambda^{-2}={}$const.

Although physically significant quantities such as $\gamma_{\phi}$ or
$\beta(g)$ are independent of $\dot{\alpha}$ the same is not true of the
vertex functions $u_{n}$. In particular one should be careful with the other
wise very convenient choice $\dot{\alpha}={}$const. For this choice of
$\dot{\alpha}$ the vertex function contain parts that are divergent as
$d\rightarrow4$. This is annoying but nothing more. The way around this
problem is the usual one, to think of the $u_{n}$'s as separated into two
parts $u_{n}=u_{n}^{R}+u_{n}^{C}$ one of which is finite while the other is a
divergent counterterm. The RG equations (\ref{ch var g}) keep track of the
evolution of both the finite and the divergent parts. The presence of these
divergences is a manifestation of the fact that while the RG was historically
connected to renormalization theory, such a connection, although sometimes
convenient, is not at all necessary.

\section{Conclusions}

In this work an approach to the renormalization group has been developed in
which the RG transformations are convenient changes of variables. The main
conclusions of this work are enumerated below.

\begin{enumerate}
\item A class of exact infinitesimal RG transformations has been proposed. The
form of the transformations is suggested quite naturally after several known
exact RG's are formulated in a conveniently simplified way. Conversely, those
exact RG's can be treated as special cases of a more general formalism.

\item The transformations are pure changes of variables (\emph{i.e.}, no
explicit integration or elimination of some degrees of freedom is required)
such that a saddle point approximations is more accurate, becoming in some
cases, asymptotically exact as the transformations are iterated.

\item Solutions of the RG equations for a scalar field theory were obtained
both as an expansion in $\varepsilon=4-d$ and as an expansion in a single
coupling constant. Physically significant results agree with those obtained
following conventional methods. The well-known fact that physical quantities
such as critical exponents are independent of the particular RG employed
emerges quite clearly.
\end{enumerate}

The consideration of RG's from this generalized point of view has a number of
attractive features which immediately suggest many possible applications. For
example, the method cold be extended to any problem where a saddle
approximation is used. One could perhaps obtain improved large $N$ expansions.

The role of dilatations is de-emphasized and one might profitably attack
problems where the issue is not the symmetry under scale transformations or
its breaking. It should be possible to study the phenomena of dynamical
symmetry breaking or of dynamical symmetry restoration. The localization of
the minima of the classical action $S$ and of the RG-evolved action $S_{\tau}$
need not coincide and it is the latter that will give more reliable
information about the true minima. Another related possible application could
be in the study of anomalies. Again, the true symmetry of a quantum theory
could be more reliably established by classically studying the RG-evolved
action $S_{\tau},,$ which includes some quantum effects, than by classically
studying the action $S$.

A further attractive feature is that the exact RG's formulated above do not
require the successive elimination of certain degrees of freedom and can
therefore be applied to systems with a small number or even just one degree of
freedom (see appendix B). They can also be exactly applied to field theories
defined on a lattice (see \cite{Caticha 1984b}).

Finally, although for simplicity we have only dealt with scalar field theories
symmetric under $\phi\rightarrow-\phi$ the extension to other fields,
fermions, etc. is straightforward. As mentioned in the Introduction the
original motivation for this study was to construct a gauge covariant RG
transformation. Once one establishes that changes of variables of the form
\[
\phi(x)\rightarrow\phi(x)-\delta\tau\,\dot{\alpha}_{\tau}(-i\partial
)\frac{\delta S_{\tau}[\phi]}{\delta\phi(x)}~,
\]
are indeed RG transformations, the problem of gauge covariance is easily
solved by replacing ordinary derivatives by covariant derivatives,
\begin{equation}
A(x)\rightarrow A(x)-\delta\tau\,\dot{\alpha}_{\tau}(-iD)\frac{\delta S_{\tau
}[A]}{\delta A(x)}~.
\end{equation}
Perhaps the simplest choice is $\dot{\alpha}_{\tau}={}$const. A detailed study
of the application of this kind of RG to non-Abelian gauge theories is the
subject of a companion article \cite{Caticha 1984b}.

\noindent\textbf{Acknowledgments:} I would like to thank Professor F.
Zachariasen and very especially N\'{e}stor Caticha for many stimulating
discussions and encouragement.

\section*{Appendix A. The Sharp-Cutoff RG}

In this appendix we integrate out the $\sigma$ fields in the thin momentum
shell and deduce eqs.(\ref{sharp cutoff a}) and (\ref{sharp cutoff b}). As
discussed in section 2.1 this process involves three steps.

\noindent\textbf{Step 1: }Expand $S_{\tau}[\phi+\sigma]$ in a power series
about $\sigma=0$,
\begin{equation}
S_{\tau}[\phi+\sigma]=\sum_{p=0}^{\infty}\frac{1}{p!}\int dx_{1}\ldots
dx_{p}\,S_{\tau}^{(p)}[\phi;x_{1}\ldots x_{p})\sigma(x_{1})\ldots\sigma
(x_{p})~, \tag{A.1}%
\end{equation}
where
\[
S_{\tau}^{(p)}[\phi;x_{1}\ldots x_{p})=\frac{\delta^{p}S_{\tau}[\phi]}%
{\delta\phi(x_{1})\ldots\delta\phi(x_{p})}\,.
\]

\noindent\textbf{Step 2: }Identify a convenient $\sigma$-field propagator.
Rewrite the quadratic term in eq.(A.1) as
\begin{align*}
\frac{1}{2}\int dx_{1}dx_{2}\,S_{\tau}^{(2)}[\phi;x_{1}x_{2})\sigma
(x_{1})\sigma(x_{2})  &  =\frac{1}{2}\int dx\,\sigma(x)\partial^{2}%
\sigma(x)+\\
&  +\frac{1}{2}\int dx_{1}dx_{2}\,\bar{S}_{\tau}^{(2)}[\phi;x_{1}x_{2})~,
\end{align*}
and let $S_{\tau}^{(p)}=\bar{S}_{\tau}^{(p)}$ for $p\neq2$, so that a
convenient propagator is
\begin{equation}
\Delta_{\tau}(x-y)=\int d\tilde{q}\,\Delta_{\tau}(q)e^{-iq(x-y)}=\frac
{\delta\tau\Lambda_{\tau}^{d-2}}{(2\pi)^{d}}\int d\Omega_{d}\,e^{-iq(x-y)}~.
\tag{A.2}%
\end{equation}
where $d\tilde{q}=d^{d}q/(2\pi)^{d}$.

\noindent\textbf{Step 3: }Treat the $\sigma^{p}$ vertices perturbatively.
Rewrite eq.(\ref{new action}) in the form
\begin{align*}
\exp-S_{\tau+\delta\tau}[\phi]  &  =\exp-\left[  \sum_{p=0}^{\infty}\frac
{1}{p!}\int dx_{1}\ldots dx_{p}\,\bar{S}_{\tau}^{(p)}\frac{\delta}{\delta
j(x_{1})}\ldots\frac{\delta}{\delta j(x_{p})}\right] \\
&  \left.  \int D\sigma\,\exp-\int dx\left(  \frac{1}{2}\sigma\Delta_{\tau
}^{-1}\sigma-j\sigma\right)  \right\vert _{j=0}~.
\end{align*}
Since each $\sigma$ propagator contributes a factor of $\delta\tau$, eq.(A.2),
while each vertex $\bar{S}_{\tau}^{(p)}$ contributes a factor $(\delta
\tau)^{0}$ it follows that to first order in $\delta\tau$ only diagrams with
one internal $\sigma$ line need be included. Therefore
\[
S_{\tau+\delta\tau}[\phi]=\int dx_{1}dx_{2}\frac{1}{2}\Delta_{\tau}%
(x_{1}-x_{2})\,\left[  S_{\tau}^{(2)}(x_{1},x_{2})-S_{\tau}^{(1)}%
(x_{1})S_{\tau}^{(1)}(x_{2})\right]  +O(\delta\tau^{2})~,
\]
where we have dropped the bars which amounts to ignoring a $\phi$-independent
constant. Rewriting this expression in momentum space leads to
eq.(\ref{sharp cutoff a}) as desired.

\section*{Appendix B. A Field Theory in Zero Dimensions}

In this appendix wee consider in more detail the application of the RG
formalism described previously to a field theory in zero dimensions for which
the partition function is an ordinary integral. This study serves to clarify
the concepts in a much simpler setting exhibiting the essence of the RG
transformation as a change of variables better suited for a semiclassical
approximation, and also to illustrate a point mentioned in section 6, namely
that these RG's are not restricted to systems with an infinite number of
degrees of freedom. First we deduce the incomplete-integration RG equation
interpreting it as a change of variables and then show that a steepest descent
approximation becomes asymptotically exact for the RG evolved action. Finally,
as a practical example we perform an RG improved perturbative calculation.

As discussed in section 2.2 \textquotedblleft incomplete
integration\textquotedblright\ is achieved through the introduction of a
constant,
\[
1=N_{\alpha}^{-1}\int_{-\infty}^{\infty}dy\,\delta_{\alpha}\left(  y-x\right)
\quad\text{where}\quad N_{\alpha}=\left(  1+4\pi\alpha\right)  ^{1/2}~
\]
into the \textquotedblleft partition function\textquotedblright,
\begin{equation}
z=\int_{-\infty}^{\infty}dy\,\exp-S(y)=N_{\alpha}^{-1}\int_{-\infty}^{\infty
}dx\,\exp-S_{\alpha}(x) \tag{B.1}%
\end{equation}
where%
\begin{equation}
\exp-S_{\alpha}(x)=\int_{-\infty}^{\infty}dy\,\delta_{\alpha}\left(
y-x\right)  \exp-S(y)~. \tag{B.2}%
\end{equation}
Using
\[
\frac{d\delta_{\alpha}(x)}{d\alpha}=\frac{2\pi\delta_{\alpha}(x)}{1+4\pi
\alpha}+\frac{d^{2}\delta_{\alpha}(x)}{d\alpha^{2}}~,
\]
eq.(B.2) can be turned into an RG differential equation,
\begin{equation}
\frac{dS_{\alpha}(x)}{d\alpha}=\frac{d^{2}S_{\alpha}(x)}{d\alpha^{2}}-\left[
\frac{dS_{\alpha}(x)}{dx}\right]  ^{2}-\frac{2\pi}{1+4\pi\alpha}~. \tag{B.3}%
\end{equation}
This evolution can be interpreted as a sequence of changes of variables.
Changing $x$ to $x^{\prime}=x+\eta(x)d\alpha$ in (B.1) leads to
\[
z=N_{\alpha+d\alpha}^{-1}\int_{-\infty}^{\infty}dx\,\exp-S_{\alpha+d\alpha
}(x)~,
\]
where
\[
S_{\alpha+d\alpha}(x)=S_{\alpha}(x)+\frac{dS_{\alpha}(x)}{dx}\eta
d\alpha-\frac{d\eta}{dx}d\alpha-\frac{2\pi d\alpha}{1+4\pi\alpha}~.
\]
If one chooses $\eta(x)=-dS_{\alpha}/dx$ this is precisely the RG equation (B.3).

Equation (B.3) can be transformed into a system of differential equations for
the evolution of the \textquotedblleft vertex functions\textquotedblright.
Substituting
\[
S_{\alpha}(x)=\sum_{n=0,\,\text{even}}\frac{1}{n!}u_{n}(\alpha)x^{n}\,,
\]
into (B.3) one obtains%
\begin{equation}
\frac{du_{0}}{d\alpha}=u_{2}-\frac{2\pi}{1+4\pi\alpha}~, \tag{B.4a}%
\end{equation}
and for $n\neq0$,
\begin{equation}
\frac{du_{n}}{d\alpha}=u_{n+2}-\sum_{m=2}^{n}\binom{n}{m-1}u_{m}u_{n-m+2}~.
\tag{B.4b}%
\end{equation}

Next we come to the question of why it is useful to go through the trouble of
solving (B.3). Consider a steepest descent approximation to (B.1):
\[
z=\frac{1}{N_{\alpha}}\left[  \frac{2\pi}{S_{\alpha}^{(2)}(\bar{x}_{\alpha}%
)}\right]  ^{1/2}\exp-S_{\alpha}(\bar{x}_{\alpha})+\ldots
\]
where $S_{\alpha}^{(n)}(\bar{x}_{\alpha})$ is the $n$-th derivative of
$S_{\alpha}$ and $\bar{x}_{\alpha}$ is the saddle point, $S_{\alpha}%
^{(1)}(\bar{x}_{\alpha})=0$. The incomplete integration was designed so that
as $\alpha\rightarrow\infty$ the exponential factor on the right hand side,
$\exp(-S_{\alpha})$, tends to the desired exact value $z$. It the leading
steepest descent approximation is to become exact it should be true that
\begin{equation}
\lim_{\alpha\rightarrow\infty}\frac{1}{N_{\alpha}}\left[  \frac{2\pi
}{S_{\alpha}^{(2)}(\bar{x}_{\alpha})}\right]  ^{1/2}=1~. \tag{B.5}%
\end{equation}
It is easy to see that this is so by referring back to eq.(B.3). As
$\alpha\rightarrow\infty$ the left hand side vanishes. Evaluating at the
saddle point $\bar{x}_{\alpha}$, the second term on the right also vanishes
and one gets,
\[
S_{\alpha}^{(2)}(\bar{x}_{\alpha})\approx\frac{2\pi}{1+4\pi\alpha}~,
\]
which implies (B.5) as desired.

It is interesting to see what is happening from another point of view.
Consider evaluating
\begin{equation}
z=\int_{-\infty}^{\infty}dx\,\exp-\left(  \frac{1}{2}wx^{2}+\frac{1}%
{4!}\lambda x^{4}\right)  ~. \tag{B.7}%
\end{equation}
If $\lambda$ is small enough one could try a perturbative expansion,
\begin{equation}
z\approx\left(  \frac{2\pi}{w}\right)  ^{1/2}\left(  1-\frac{\lambda}{8w^{2}%
}+O(\lambda^{2})\right)  ~. \tag{B.8}%
\end{equation}
But one could refer to eq.(B.1) and try an RG-improved expansion,
\begin{equation}
z\approx\frac{1}{N_{\alpha}}e^{u_{0}(\alpha)}\left(  \frac{2\pi}{u_{2}%
(\alpha)}\right)  ^{1/2}\left(  1-\frac{1}{8}\frac{u_{4}(\alpha)}{u_{2}%
^{2}(\alpha)}+\ldots\right)  ~, \tag{B.9}%
\end{equation}
where $u_{0}(\alpha)$, $u_{2}(\alpha)$, and $u_{4}(\alpha)$ are solutions of
(B.4) with the initial conditions $u_{0}(0)=0$, $u_{2}(0)=w$, and
$u_{4}(0)=\lambda$. To order $\lambda$ these solutions are
\begin{align}
u_{0}(\alpha)  &  =\frac{1}{2}\log\frac{1+2\pi\alpha}{1+4\pi\alpha}%
+\frac{\lambda\alpha^{2}}{2(1+2w\alpha)^{2}}+O(\lambda^{2})\,,\tag{B.10a}\\
u_{2}(\alpha)  &  =\frac{\omega}{1+2w\alpha}+\frac{\lambda\alpha}%
{(1+2w\alpha)^{3}}+O(\lambda^{2})\,,\tag{B.10b}\\
u_{4}(\alpha)  &  =\frac{\lambda}{(1+2w\alpha)^{4}}+O(\lambda^{2})\,.
\tag{B.10c}%
\end{align}
According to (B.10b) $u_{2}(\alpha)$ tends to vanish as $\alpha$ increases.
This could mean trouble since the integrand becomes flatter and flatter (this
is apparent in eq.(B.6) also). The reliability of the steepest descent
approximation would become increasingly doubtful. Fortunately we are saved by
(B.10c) which shows that the \textquotedblleft interaction\textquotedblright%
\ $u_{4}$ vanishes much faster and the integrand approaches a Gaussian, a
rather flat one but still a Gaussian. The perturbative correction $u_{4}%
/u_{2}^{2}$ in eq.(B.9) asymptotically vanishes.

Given the vertices (B.10) correct to $O(\lambda)$ the best approximation is
obtained letting $\alpha\rightarrow\infty$, which gives
\begin{equation}
z\approx\left(  \frac{2\pi}{w}\right)  ^{1/2}\exp-\frac{\lambda}{8w^{2}}\,,
\tag{B.11}%
\end{equation}
which has the typical exponential of RG improved calculations.

It is quite remarkable that one can exhibit the powerful RG techniques in such
a simple example. it is perhaps even more remarkable that in this simple study
even their limitations become apparent. For the toy model (B.7) the exact
result is known and for strong coupling (large $\lambda$) the correct
dependence on $\lambda$ is the power law
\[
z\approx\frac{1}{2}\Gamma(\frac{1}{4})\left(  \frac{6}{\lambda}\right)
^{1/4}~,
\]
and not the exponential dependence of (B.11). This illustrates the known fact
that while RG perturbation expansions are an improvement over plain
perturbation expansions, they remain nevertheless restricted to the small
coupling regime. Needless to say, this is not a limitation of the RG itself
(eqs.(B.3-4) are exact) but of the perturbative solution (B.10) to the RG
equation (B.4).

\section*{Appendix C. The $\varepsilon$ Expansion}

The RG system of equations (\ref{ch var g}) is greatly simplified if one
realizes that the solutions of interest are not the most general solutions of
the first order differential equations in which the momenta $q_{j}$ are
independent variables, but rather those special solutions with interesting
scaling properties when all $q_{j}$'s are scaled together. The unwanted
solutions can be discarded by evaluating (\ref{ch var g}) at momenta $\lambda
q_{j}$ with $\lambda=e^{\tau}$ instead of $q_{j}$. This eliminates the partial
derivatives and (\ref{ch var g}) becomes%
\begin{gather}
\left\{  \lambda\frac{d}{d\lambda}+nd_{\phi}-d\right\}  u_{n}(\lambda
q_{1}\ldots\lambda q_{n},\lambda)=\int d\tilde{k}\,\dot{\alpha}(k)u_{n+2}%
(\lambda q_{1}\ldots\lambda q_{n},k,-k,\lambda)+\nonumber\\
\sum_{m\geq2}^{n}\binom{n}{m-1}\frac{1}{n!}\sum_{\{q_{j}\}}\dot{\alpha
}(\lambda k_{m})u_{m}(\lambda q_{1}\ldots\lambda q_{m-1},\lambda k_{m}%
,\lambda)\nonumber\\
\times u_{n-m+2}(\lambda q_{m}\ldots\lambda q_{n},-\lambda k_{m},\lambda)
\tag{C.1}%
\end{gather}
Substituting eq.(\ref{sol 1a}) and (\ref{sol 1b}) into (C.1) leads to a set of
first order ordinary differential equations for the $V$'s,
\begin{align}
\left(  \lambda\frac{d}{d\lambda}-2\right)  V_{20}  &  =-2\dot{\alpha}%
V_{20}^{2}~,\tag{C.2}\\
\left(  \lambda\frac{d}{d\lambda}-2\right)  V_{21}+2\gamma_{1}V_{20}  &  =\int
d\tilde{k}\,\dot{\alpha}V_{41}-4\dot{\alpha}V_{20}V_{21}~,\tag{C.3}\\
\left(  \lambda\frac{d}{d\lambda}-2\right)  V_{22}+2\gamma_{1}V_{21}%
+2\gamma_{2}V_{20}  &  =\int d\tilde{k}\,\dot{\alpha}V_{42}+\frac
{d}{d\varepsilon}\left.  \int d\tilde{k}\,\dot{\alpha}V_{41}\right\vert
_{\varepsilon=0}\nonumber\\
&  -2\dot{\alpha}V_{21}^{2}-4\dot{\alpha}V_{20}V_{22} \tag{C.4}%
\end{align}

\begin{align}
\lambda\frac{d}{d\lambda}V_{41}  &  =-2(%
%TCIMACRO{\tsum \limits_{j}^{4}}%
%BeginExpansion
{\textstyle\sum\limits_{j}^{4}}
%EndExpansion
\dot{\alpha}V_{20})V_{41}~,\tag{C.5}\\
\lambda\frac{d}{d\lambda}V_{42}+(4\gamma_{1}-1)V_{41}  &  =\int d\tilde
{k}\,\dot{\alpha}V_{62}-2(%
%TCIMACRO{\tsum \limits_{j}^{4}}%
%BeginExpansion
{\textstyle\sum\limits_{j}^{4}}
%EndExpansion
\dot{\alpha}V_{20})V_{42}-2(%
%TCIMACRO{\tsum \limits_{j}^{4}}%
%BeginExpansion
{\textstyle\sum\limits_{j}^{4}}
%EndExpansion
\dot{\alpha}V_{21})V_{41}~,\tag{C.6}\\
\left(  \lambda\frac{d}{d\lambda}-2\right)  V_{62}  &  =-2(%
%TCIMACRO{\tsum \limits_{j}^{6}}%
%BeginExpansion
{\textstyle\sum\limits_{j}^{6}}
%EndExpansion
\dot{\alpha}V_{20})V_{62}-2(\dot{\alpha}V_{41}V_{41}+{}\text{perm.})~.
\tag{C.7}%
\end{align}
The arguments of the $V$'s can be easily obtained by referring back to
eq.(C.1). In eq.(C.7) the ten permutations refer to the inequivalent ways of
grouping six momenta into two sets of three.

Equation (C.2) is of the Bernoulli type. The solution behaving as
$V_{20}\approx q^{2}$ for small $q$ is
\[
V_{20}(q)=q^{2}f(q)~,
\]
where
\begin{equation}
f(q)=\frac{1}{1+q^{2}\int_{0}^{1}d\lambda^{2}\dot{\alpha}(\lambda q)}%
=\exp-2\int_{0}^{1}\frac{d\lambda}{\lambda}\dot{\alpha}(\lambda q)V_{20}%
(\lambda q)~. \tag{C.8}%
\end{equation}
The second equality is very useful because it will allow us to construct
integrating factors for all the remaining eqs.(C3-7) which are linear. The
solution to (C.5) is
\[
V_{41}(q_{1}\ldots q_{4})=A\prod_{j=1}^{4}f(q_{j})~.
\]
Next we solve (C.3). The fixed point ($\partial V_{21}/\partial\lambda=0$) and
the analyticity requirements force us to choose $\gamma_{1}=0$ so that
\[
V_{20}(q)=(Cq^{2}-\frac{1}{2}AB)f^{2}(q)~,
\]
where $C$ is a constant and
\begin{equation}
B=\int d\tilde{k}\,\dot{\alpha}(k)f^{2}(k)~. \tag{C.9}%
\end{equation}
This completes the solution to order $\varepsilon$.

The solution for (C.7) is straightforward,
\begin{equation}
V_{62}(q_{1}\ldots q_{4})=-A^{2}\left[  \prod_{j=1}^{4}f(q_{j})\right]
\left[  D(q_{1}+q_{2}+q_{3})+{}\text{perm.}\right]  ~, \tag{C.10}%
\end{equation}
where
\begin{equation}
D(k)=\frac{1}{k^{2}}\left[  1-f(k)\right]  ~.
\end{equation}
The solution for $V_{42}$ is messier; the only important point being that in
order to eliminate a divergence at $\lambda=0$ (or $q=0$) the constant $A$ in
$V_{41}$ must be chosen to be $A=(4\pi)^{2}/3$.

Finally, we turn to $V_{22}$. Again its explicit form is not very illuminating
but the requirement that it be analytic at $q=0$ determines both
$V_{42}(0\ldots0)$ which is not in itself very interesting and also
$\gamma_{2}$,
\begin{equation}
\gamma_{2}=-\frac{(4\pi)^{4}}{18}\int d\tilde{k}d\tilde{k}^{\prime}\frac
{1}{k^{2}}f(k)f^{\prime}(k^{\prime})f^{\prime\prime}(Q)~, \tag{C.11}%
\end{equation}
where $\vec{Q}=\vec{k}+\vec{k}^{\prime}$ and $f^{\prime}=df(k)/dk^{2}$. The
integral in eq.(C.11) can be done analytically for $\dot{\alpha}=\Lambda
^{-2}\exp(q^{2}/\Lambda^{2})$, or else numerically for $\dot{\alpha}%
=\Lambda^{-2}$ or for $\dot{\alpha}=q^{2}/\Lambda^{4}$. The result is
$\gamma_{2}=1/108$ so that
\[
\gamma_{\phi}=\frac{\varepsilon^{2}}{108}\
\]
as it should be \cite{Wilson Kogut 1974}. This completes the solution to order
$\varepsilon^{2}$.

\section*{Appendix D. The Perturbative Solution}

Substituting eq.(\ref{sol 2a}) and (\ref{sol 2b}) into (C.1) leads to a set of
first order ordinary differential equations for the $U$'s,
\begin{align}
\left(  \lambda\frac{d}{d\lambda}-2\right)  U_{20}  &  =-2\dot{\alpha}%
U_{20}^{2}~,\tag{D.1}\\
\left(  \lambda\frac{d}{d\lambda}-2+b_{1}\right)  U_{21}+2\gamma_{1}U_{20}  &
=\Lambda^{\varepsilon}\int d\tilde{k}\,\dot{\alpha}U_{41}\nonumber\\
&  -4\dot{\alpha}U_{20}U_{21}~,\tag{D.2}\\
\left(  \lambda\frac{d}{d\lambda}-2+2b_{1}\right)  U_{22}+(b_{2}+2\gamma
_{1})U_{21}+2\gamma_{2}U_{20}  &  =\Lambda^{\varepsilon}\int d\tilde{k}%
\,\dot{\alpha}U_{42}\nonumber\\
&  -2\dot{\alpha}U_{21}^{2}-4\dot{\alpha}U_{20}U_{22} \tag{D.3}%
\end{align}

\begin{align}
\left(  \lambda\frac{d}{d\lambda}+b_{1}-\varepsilon\right)  U_{41}  &  =-2(%
%TCIMACRO{\tsum \limits_{j}^{4}}%
%BeginExpansion
{\textstyle\sum\limits_{j}^{4}}
%EndExpansion
\dot{\alpha}U_{20})U_{41}~,\tag{D.4}\\
\left(  \lambda\frac{d}{d\lambda}+2b_{1}-\varepsilon\right)  U_{42}%
+(b_{2}+4\gamma_{1})U_{41}  &  =\Lambda^{\varepsilon}\int d\tilde{k}%
\,\dot{\alpha}U_{62}-2(%
%TCIMACRO{\tsum \limits_{j}^{4}}%
%BeginExpansion
{\textstyle\sum\limits_{j}^{4}}
%EndExpansion
\dot{\alpha}U_{20})U_{42}\nonumber\\
&  -2(%
%TCIMACRO{\tsum \limits_{j}^{4}}%
%BeginExpansion
{\textstyle\sum\limits_{j}^{4}}
%EndExpansion
\dot{\alpha}U_{21})U_{41}~,\tag{D.5}\\
\left(  \lambda\frac{d}{d\lambda}+2b_{1}+2-2\varepsilon\right)  U_{62}  &
=-2(%
%TCIMACRO{\tsum \limits_{j}^{6}}%
%BeginExpansion
{\textstyle\sum\limits_{j}^{6}}
%EndExpansion
\dot{\alpha}U_{20})U_{62}-2(\dot{\alpha}U_{41}U_{41}+{}\text{perm.})~.
\tag{D.6}%
\end{align}
These equations are naturally very similar to those of appendix C and the
solution proceeds exactly as before except that now, as discussed in section
5, no requirement of analyticity is made.

The solution of (D.1) behaving as $U_{20}\approx q^{2}$ for small $q$ is
\[
U_{20}(q)=q^{2}f(q)
\]
with $f(q)$ given by (C.8) as before. The integration of (D.4) is
straightforward.\ We want $U_{41}$ independent of $\lambda$ so that all the
evolution of $gU_{41}$ is attributed to the evolution of $g$. This forces us
to choose $b_{1}=\varepsilon$. Further we normalize $U_{41}(0\ldots0)=1$ which
is the conventional normalization for $g$. Then
\[
U_{41}(q_{1}\ldots q_{4})=\prod_{j=1}^{4}f(q_{j})~.
\]
Next consider (D.2). The requirement that $U_{21}$ is independent of $\lambda$
implies $\gamma_{1}=0$ and one obtains the generally non-analytic expression
\[
U_{21}(q)=\Lambda^{\varepsilon}\left(  Cq^{2-\varepsilon}-\frac{B}%
{2-\varepsilon}\right)  f^{2}(q)~,
\]
where $B$ is given by (C.9). The solution of $U_{62}$ is uneventful; one
obtains the expression (C.10) with $A=1$. Finally, the solution of (D.5) for
$U_{42}$ does not depend on $\lambda$ provided one chooses $b_{2}%
=-3/(4\pi)^{2}+O(\varepsilon)$, that is
\[
-\frac{dg}{d\tau}=\beta(g)=-\varepsilon g+\left(  \frac{3}{(4\pi)^{2}%
}+O(\varepsilon)\right)  g^{2}+\ldots
\]
as expected. We stop here since the solution for $U_{22}$ and $\gamma_{2}$
proceeds in the same way.

One final comment concerning the choice of \textquotedblleft
cutoff\textquotedblright\ function $\dot{\alpha}$: if one chooses a constant
$\dot{\alpha}=\Lambda^{-2}$ the vertex functions develop divergences as
$d\rightarrow4$. This is evident when one computes the constant $B$ given by
(C.9). As discussed further in section 5 this is not really a problem,
particularly since physically significant quantities such as $\gamma_{\phi
}(g)$ and $\beta(g)$ are perfectly finite and independent of $\dot{\alpha}$.


\begin{thebibliography}{99}                                                                                               %


\bibitem {Caticha 1984a}A. Caticha, Caltech preprint CALT-68-1099 (1984); and
also in \emph{Changes of Variables and the Renormalization Group}, Ph.D.
Thesis, California Institute of Technology (1985).

\bibitem {Wilson Kogut 1974}K. Wilson and J. Kogut, Phys. Rep. \textbf{C12},
75 (1974).

\bibitem {Wilson 1983}K. Wilson, Rev. Mod. Phys. \textbf{55}, 583 (1983).

\bibitem {Gell-Mann Low 1954}M. Gell-Mann and F. Low, Phys. Rev. \textbf{95},
1300 (1954).

\bibitem {Jona-Lasinio 1973}G. Jona-Lasinio, Proc. Nobel Symposium, Vol.
\textbf{24}, 38 (New York, NY, 1973).

\bibitem {Wegner 1974}F. Wegner, J. Phys. \textbf{C7}, 2098 (1974).

\bibitem {Wegner Houghton 1973}F. Wegner and A. Houghton, Phys. Rev.
\textbf{A8}, 401 (1973).

\bibitem {Wilson 1976}K. Wilson, p. 243 in \emph{New Pathways in High Energy
Physics}, ed. by A. Perlmutter (Plenum, New York 1976).

\bibitem {Lowenstein Mitter 1977}J. H. Lowenstein and P. K. Mitter, Ann. Phys.
\textbf{105}, 138 (1977).

\bibitem {Mitter Valent 1977}P. K. Mitter and G. Valent, Phys. Lett.
\textbf{B70}, 65 (1977).

\bibitem {Shalloway 1979}D. Shalloway, Phys. Rev. \textbf{D19}, 1762 (1979).

\bibitem {Baker Ball Zachariasen  1983}M. Baker, J. S. Ball, and F.
Zachariasen, Nucl. Phys. \textbf{B229}, 445 (1983).

\bibitem {Caticha 1984b}A. Caticha, \textquotedblleft A Gauge Covariant
Renormalization Group,\textquotedblright\ Caltech preprint, CALT-68-1222 (1985).

\bibitem {Gervais Jevicki 1976}J.-L. Gervais and A. Jevicki, Nucl. Phys.
\textbf{B110}, 93 (1976).

\bibitem {Fujikawa 1979}K. Fujikawa, Phys. Rev. Lett. \textbf{42}, 1195 (1979).

\bibitem {Shukla Green 1974}P. Shukla and M. Green, Phys. Rev. Lett.
\textbf{33}, 1263 (1974).

\bibitem {Polchinski 1984}J. Polchinski, Nucl. Phys. \textbf{B231}, 269 (1984).

\bibitem {Wetterich 1991}C. Wetterich, Nucl. Phys. \textbf{B352}, 529 (1991);
Phys. Lett. \textbf{B301}, 90 (1993).

\bibitem {Reuter Wetterich 1994}M. Reuter and C. Wetterich, Nucl. Phys.
\textbf{B417}, 181 (1994).

\bibitem {Reuter 1998}M. Reuter, Phys. Rev. \textbf{D57}, 971 (1998).

\bibitem {Latorre Morris 2000}J. I. Latorre and T. R. Morris, J. High Energy
Phys. \textbf{11}, 004 (2000); arXiv:hep-th/0008123.

\bibitem {Morris Preston 2016}T. R. Morris and A. W. H. Preston,
\textquotedblleft Manifestly diffeomorphism invariant classical Exact
Renormalization Group,\textquotedblright\ arXiv:1602.08993.

\bibitem {Morris 1994}T. R. Morris, Int. J. Mod. Phys. \textbf{A9}, 2411
(1994); arXiv:hep-ph/9308265.

\bibitem {Berges Tetradis Wetterich  2000}J. Berges, N. Tetradis, and C.
Wetterich, Phys.Rep. \textbf{363}, 223 (2002); arXiv: hep-th/0005122.

\bibitem {Delamotte 2007}B. Delamotte, \textquotedblleft An introduction to
the nonperturbative renormalization group,\textquotedblright\ arXiv:cond-mat/0702365.

\bibitem {Reuter Saueressig 2007}M. Reuter and F. Saueressig,
\textquotedblleft Functional Renormalization Group Equations, Asymptotic
Safety, and Quantum Einstein Gravity,\textquotedblright\ arXiv:0708.1317.

\bibitem {Zinn-Justin 2007}J. Zinn-Justin, \emph{Phase Transitions and the
Renormalization Group} (Oxford U.P., 2007).

\bibitem {Rosten 2010}O.J. Rosten, \textquotedblleft Fundamentals of the Exact
renormalization Group,\textquotedblright\ Phys.Rep. \textbf{511}, 177 (2012); arXiv:1003.1366.

\bibitem {Nagy 2012}S. Nagy, \textquotedblleft Lectures on renormalization and
asymptotic safety,\textquotedblright\ arXiv:1211.4151.
\end{thebibliography}
\end{document}